\documentclass[iop]{emulateapj}		% Nice compact 2-col format for ApJ

\newcommand{\chisq}{\mbox{$\chi^2$}}		% Chi squared
\newcommand{\hst}{{\it HST}}			% Abbrev. for Hubble Space Tel.
\newcommand{\ks}{\mbox{$K_s$}}			% Passband Ks
\newcommand{\kslim}{\mbox{$K_{s,{\rm lim}}$}}	% Ks 50% completeness limit
\newcommand{\kslimfnd}{\mbox{$K_{s,{\rm lim, fnd}}$}} % Same f/clusters found
\newcommand{\kslimacc}{\mbox{$K_{s,{\rm lim, acc}}$}} % Same f/clusters acceptd
\newcommand{\n}{NGC~}				% NGC = New General Catalogue
\newcommand{\ubv}{\mbox{$U\!BV$}}		% UBV colors
\newcommand{\vi}{\mbox{$V\!-\!I$}}              % Color index V-I
\newcommand{\viks}{\mbox{$V\!IK_s$}}		% VIKs passbands
\newcommand{\vks}{\mbox{$V\!-\!K_s$}}		% V-Ks
\usepackage{subfigure}

\shorttitle{Globular Clusters in Four Merger Remnants}
\shortauthors{Trancho et al.}

\begin{document}

\title{Intermediate-age Globular Clusters in Four Galaxy Merger Remnants}  


\author{Gelys Trancho}
\affil{Giant Magellan Telescope Organization, 251 S. Lake Avenue, Pasadena, CA 91101, USA,}
\affil{Carnegie Observatories, 813 Santa Barbara Street, Pasadena, CA 91101, USA}
\email{gtrancho@gmto.org}
\author{Bryan W.\ Miller}
\affil{Gemini Observatory, Casilla 603, La Serena, Chile}
\author{Fran\c{c}ois Schweizer}
\affil{Carnegie Observatories, 813 Santa Barbara Street,  Pasadena, CA 91101, USA}
\author{Daniel P.\ Burdett}
\affil{The University of Adelaide, South Australia 5005, Australia}
\and
\author{David Palamara}
\affil{Monash University, Clayton, Victoria 3800, Australia}


\begin{abstract}

We present the results of combining {\em Hubble Space Telescope} optical
photometry with ground-based \ks-band photometry from the Gemini imagers
NIRI and FLAMINGOS-I to study the globular-cluster populations in four
early-type galaxies that are candidate remnants of recent mergers (\n1700,
\n2865, \n4382, and \n7727).
These galaxies were chosen based on their blue colors and fine structure,
such as shells and ripples that are indicative of past interactions.
We fit the combined \viks\ globular-cluster data with simple toy models of
mixed cluster populations that contain three subpopulations of different
age and metallicity.
The fits, done via Chi-square mapping of the parameter space, yield clear
evidence for the presence of intermediate-age clusters in each galaxy.
We find that the ages of $\sim$\,1\,--\,2 Gyr for these globular-cluster
subpopulations are consistent with the previously estimated merger ages for
the host galaxies.

\end{abstract}


\section{INTRODUCTION}
Studies of globular cluster (GC) systems play a critical role in our
understanding of galaxy formation. Until two decades ago it was thought that
all GCs were old, and this shaped our view of how galaxies, especially
elliptical galaxies, formed (see \citealt{harris91}).  Imaging with the
{\em Hubble Space Telescope (HST)} then revealed that young compact star
clusters form copiously in galaxy mergers \citep{whitmore93,miller97,zepf99},
a fact that supports theories in which giant elliptical galaxies are formed
through mergers of spirals \citep{schweizer87,ashman92}.

However, the formation and evolution of GC systems is still not well
understood.  We should be able to observe how GC systems evolve from the
very young systems with power-law luminosity functions into old systems
with log-normal luminosity functions, such as are observed in typical
elliptical galaxies.  These observations would greatly clarify exactly which
processes (shocking, evaporation, dynamical friction; e.g., \citealt{fall01})
are important in the evolution of star cluster systems.

Finding intermediate-age GC populations in elliptical galaxies has been
difficult.  Spectroscopy is often thought to be a relatively accurate method
for measuring cluster ages and metallicities.
However, in practice spectroscopy is very time consuming, even for
8\,m-class telescopes, leaving only the brightest GCs open to such
analyses.  \citet{strader03} spent 9.3 hours observing nine GC candidates
in the E5 galaxy \n3610 with Keck, and found only one young GC.

The majority of candidate GCs in galaxies identified via morphology and
color to be intermediate-age merger remnants are too faint for spectroscopic
study.
Photometric studies have also had limited success, partly because most
have been done only at optical wavelengths (specifically, in the $V$ and $I$
passbands). The problem is that the \vi\ colors of metal-rich GCs with
ages of 1--3 Gyr overlap with the colors of $\sim$13 Gyr old metal-poor
GCs, resulting in an age-metallicity degeneracy
\citep[esp.\ Fig.\ 15]{whitmore97}.
Fortunately, the combination of optical and near-infrared (NIR)
photometry can break this age--metallicity degeneracy.  This is because
optical-to-NIR indices are sensitive to metallicity, but only weakly
dependent on age \citep[e.g.,][]{kissler00,puzia02}.
{ Physically
this works because in older stellar populations the $V$-band predominantly
samples the light of stars near the turn-off, while the \ks-band is more
sensitive to cooler stars on the giant branch.}
The color of the { giant branch}
is metallicity-dependent, so the index \vks\ is primarily a metallicity
indicator.  Thus, we can lift the age--metallicity degeneracy by plotting
the locations of Simple Stellar Populations (SSP) in a color--color diagram, 
specifically  \vi\ vs.\ \vks\,.

\section{OBSERVATIONS AND REDUCTIONS}\label{sec:obs}

\subsection{Sample Selection} 

The four early-type galaxies \n1700, \n2865, \n4382, and \n7727 were
selected based on their tidal fine-structure content, blue $U\!BV$ colors,
and enhanced Balmer absorption lines \citep{schweizer92,schweizer90}.
The sample spans the morphology and age gap between on-going mergers and
normal elliptical galaxies.

\n4382 is similar to the merger remnants \n7252, \n3921, and \n7727, but
lacks tidal tails.
Its rich fine structure and distortions (see Fig.~1 in \citealt{schweizer88}
and Fig.~9 in \citealt{kormendy09}) suggest that it experienced a merger
1\,--\,2 Gyr ago, a hypothesis strengthened by its still bluish center
\citep{fisher96}, double-peaked nucleus \citep{lauer05}, and counter-rotating
core \citep{mcdermid04}.
It was selected specifically to bridge the age gap between ongoing mergers
like \n4038/39 or \n3256 and normal old ellipticals.

The elliptical galaxies \n1700 and \n2865 both show evidence for recent
interactions or mergers and are therefore candidate ``young ellipticals.''

\n1700 features boxy isophotes, a strong isophotal twist, and two broad arms
or tails that have long suggested it may be the remnant of a major merger of
two disk galaxies (\citealt{seitzer90}, esp.\ Fig.~1b; \citealt{brown00}).
Its kinematics supports this hypothesis: it rotates fast \citep{statler96},
yet features a counterrotating core with a distinctly younger stellar
population \citep{kleineberg11}.
\citet{whitmore97} found that its GCs are consistent with an age of
$\sim\,$4 Gyr, but do not necessarily form a population of obviously
intermediate age.

\n2865 features bright irregular shells, a luminous plume of material, and
a faint outer loop, all suggesting that it is the remnant of a relatively
recent merger (see Fig.~5 of \citealt{malin83} and Fig.~1 of \citealt{hau99}).
This hypothesis is supported by \ion{H}{1} gas fragments associated with the
optical fine structure \citep{schiminovich95}, a rapidly rotating small
central core with a younger stellar population and gas \citep{hau99,serra10},
and a body with unusually blue \ubv\ colors.
\citet{sikkema06} note that in a color histogram the galaxy's GCs seem to
contain a subpopulation bluer than normal metal-poor GCs and, hence, likely
of much younger age ($\sim\,$0.5\,--\,1 Gyr).

Finally, \n7727 (= VV67 = Arp 222) is so obviously tidally disturbed that it
was included by both \citet{vorontsov59} and \citet{arp66} in their catalogs
of interacting galaxies.
Because of its spiral-shaped tidal tails it was originally classified as a
peculiar Sa galaxy \citep{devaucouleurs64,sandage81}, but was then
reclassified as a recent merger remnant of age $\sim\,$1.3 Gyr by
\citet{georgakakis00}.
The presence of bluish young star clusters in it was noted by
\citet{crabtree94} and described in some more detail by \citet{trancho04}.

\subsection{Observations}

\begin{figure}
  \begin{center}
    \includegraphics[width=0.48\textwidth]{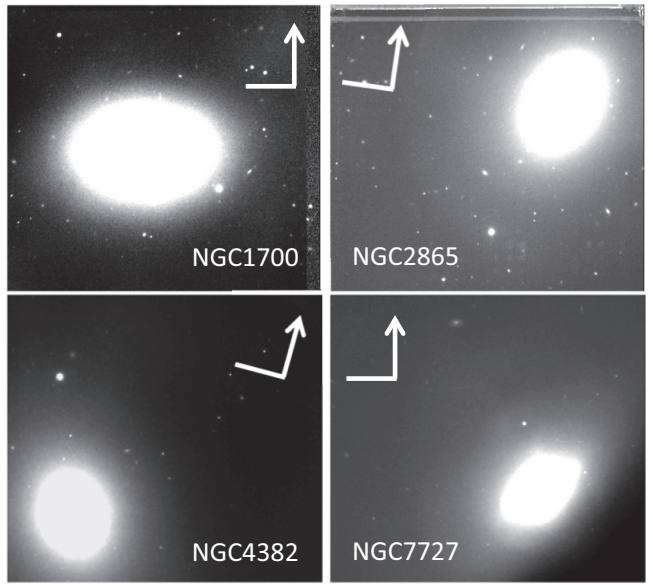}
    \caption{\ks-band images of \n1700, \n2865, and \n4382 obtained with NIRI,
             and of \n7727 obtained with FLAMINGOS-I.  The integrated exposure
             times were 0.5, 0.2, 0.5, and 1.0 hours, respectively.}
    \label{fig:NGC}
  \end{center}
\end{figure}

We obtained \ks-band images of the above four early-type galaxies
with the Near-InfraRed Imager and Spectrometer (NIRI)
on Gemini North \citep{hodapp03} and with FLAMINGOS-I on Gemini South
\citep{elston03}, with the results shown in Figure~\ref{fig:NGC}.

The program IDs were GN-2006B-Q-97 and GS-2002A-DD-4, respectively.
We used NIRI in f/6 imaging mode, which yields an image scale of
$0\farcs117$/pixel and a field of view of $2\arcmin\times 2\arcmin$, and
FLAMINGOS-I also in imaging mode, which yields a scale of $0\farcs078$/pixel
and a field of view of $2\farcm2\times 2\farcm2$.
Both instrument have similar fields of view, comparable to those of the WFPC2
and ACS cameras on \hst.

All four galaxies were also observed with \hst\, in $V$ and $I$ (GO-7468, PI:
Schweizer), and we used unpublished photometry from that project.
In the case of \n4382, we used the most recent photometric data published by
\citet{jordan09}.

Table~\ref{table0a} summarizes our observations and the data
from other authors used in this study.

%%%%%%%%%%%%%%%%%%%%%%%  Table_1  %%%%%%%%%%%%%%%%%%%%%%%%

\begin{deluxetable*}{lcccc}
\tablecolumns{5}
\tablewidth{0pt}
\tablecaption{Summary of galaxy observations and data}
\tablehead{
  \colhead{Galaxy}&
  \colhead{Distance\tablenotemark{a}}&
  \colhead{Filter}&
  \colhead{Telescope/Instrument}&
    \colhead{Author}
}
\startdata
\n1700 & 52& $V,I$ & \hst/WFPC2    & \citet{whitmore97}  \\
       &   &   \ks & Gemini/NIRI   & This Paper          \\
\n2865 & 35& $V,I$ & \hst/WFPC2    & Schweizer (Cycle 7)\tablenotemark{b} \\
       &   &   \ks & Gemini/NIRI   & This Paper          \\
\n4382 & 16& $V,I$\tablenotemark{c}& \hst/ACS & \citet{jordan09}\\
       &   &   \ks & Gemini/NIRI   & This Paper          \\
\n7727 & 26& $V,I$ & HST/WFPC2     & \citet{trancho04}   \\
       &   &   \ks & Gemini/FLAMINGOS-I&  This Paper 
\enddata
\tablenotetext{a}{Distance (in Mpc) based on recession velocity relative to the
	Local Group and $H_0 = 73$ km s$^{-1}$ Mpc$^{-1}$ \citep{freedman10}.}
\tablenotetext{b}{Unpublished data.}
\tablenotetext{c}{The data were originally in the passbands $g,z$ and
	were transformed to $V,I$ using equations by \citet[Eqs.\ (1)
	and (2)]{peng06}.}
\label{table0a}
\end{deluxetable*}
%%%%%%%%%%%%%%%%%%%%%%%%%%%%%%%%%%%%%%%%%%%%%%%%%%%%%%%%%%

\subsection{Data Reduction}

The data were reduced with the Gemini/NIRI package of tasks within IRAF.
Normalized flats were constructed from images taken with the Gemini
calibration unit.
Flat-field images with the IR lamp on and off allowed separation of the
instrumental thermal signature from the sensitivity response.

The near-IR sky level and structure varies on time scales of a few minutes.
To account for such variations we used a dither pattern in our imaging
sequences, which provided blank fields close enough in time to derive the
sky levels reliably.

Some of the NIRI frames contain an electronic pattern visible as vertical
striping with a period of eight columns and not always present in all
quadrants.  Before including the affected frames in our image processing
list we used the stand-alone python routine {\em nirinoise.py} which almost
perfectly removed the striping.  The same image reduction steps were
applied to the standard star observations as well. 

All reduced science and standard star exposures of the corresponding nights
were registered to a common coordinate system with the task {\em imcoadd},
based on geometric solutions obtained with the task {\em geomap} for NIRI.
The final \ks\ science image for each galaxy was derived by averaging
individual images, scaled to the galaxy image from the most photometric
night, with the IRAF task {\em imcombine}.

\section{PHOTOMETRY}

\subsection{Source Detection}

\begin{figure*}
  \begin{center}
    \includegraphics[angle=0,width=1\textwidth]{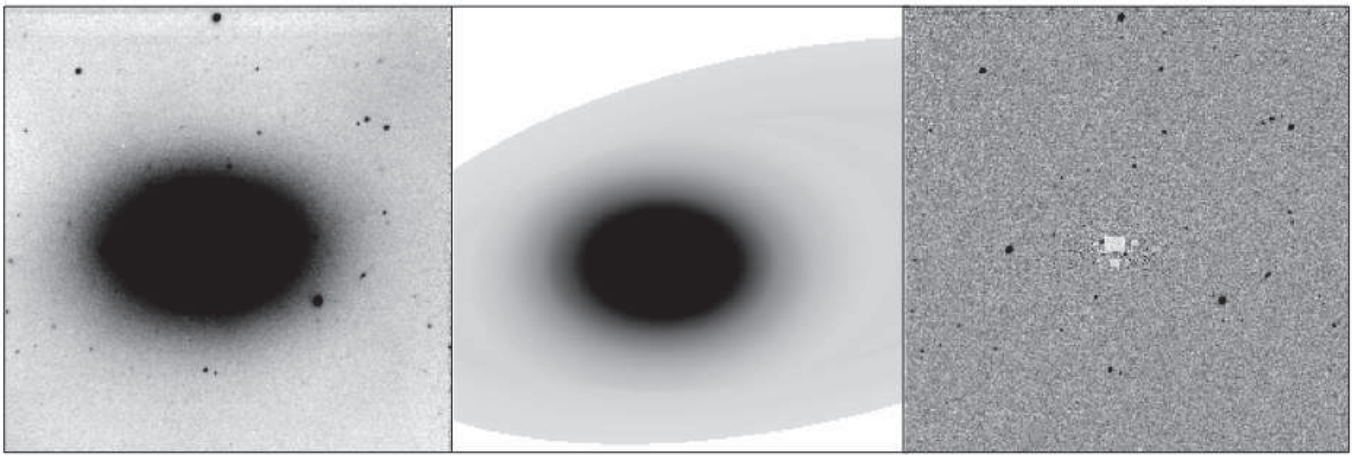}
    \caption{From left to right: \ks-band image of \n1700, galaxy model of
	     \n1700, and model-subtracted image.}
    \label{fig:technique}
  \end{center}
\end{figure*}

To detect as many GCs as possible from our \ks-band images and match them with
existing \hst\, photometry, we first created a model image of each galaxy with
the tasks {\em ellipse} and {\em bmodel} of the {\em isophote} package in IRAF. 
This model image was then subtracted from the original combined galaxy image.
This procedure facilitated the uniform detection of point sources at a
detection threshold of 3$\sigma$ above the background, using the task
{\em daofind} of the {\em daophot} package. Figure~\ref{fig:technique} shows
an example of this technique.

\subsection{Photometric Calibration}

The photometry of the candidate GCs in our sample of merger remnants
was calibrated to \ks\ on the standard system via 14\arcsec-aperture
photometry matching that used in the 2MASS survey.  This method was chosen
because the standard stars observed with Gemini indicated significant
variability in the zero point during our observations.  The conditions that
these galaxies were observed under were IQ70 and, therefore, not guaranteed to
be photometric.  The FLAMINGOS-I observations, on the other hand, showed a
high level of agreement with the calibration via the 2MASS stars;
coincidently, this night had very good photometric conditions.

\subsection{GC Selection}

We matched the optical \hst\, observations (already reduced and calibrated)
with our infrared GC candidates using the IRAF tasks {\em geomap} and
{\em geotran}.
The brightest common objects in both images served as points of reference.

\subsection{Completeness Correction}

The initial selection of GC candidates was largely based on applying
color limits to the objects identified in the $V$, $I$, and \ks\ images.
However, it was expected that this method alone would prove insufficient
to remove contaminants such as background galaxies and foreground stars.
Therefore, the photometric completeness of the data was determined, and 
a visual inspection of each GC candidate was performed. 

Since this study utilized space-based $V$ and $I$ images, but ground-based
\ks\ images, it was expected that the \ks\ images would set the completeness 
limit for the photometric data.  Hence, a set of completeness curves were
generated from the \ks\ images obtained with NIRI and FLAMINGOS-I for each
galaxy.  These curves were created through the addition of artificial
cluster images to each galaxy image, using the DAOPHOT task {\em addstar}.
For each of the galaxies \n1700, \n2865, and \n7727, 100,000 artificial
cluster images ranging from 15 to 25 magnitudes were added, with no more than
100 clusters added to any one image to prevent overcrowding.  In the case
of \n4382, which exhibits an already relatively crowded field of point
sources, only 50 artificial clusters were added to each image, effectively
halving the total number of artificial clusters studied. The DAOPHOT tasks
{\em daofind} and {\em phot} were then used to attempt to recover these
artificial clusters, and to perform photometry on them.  This process was
repeated for background values of 25, 65, 85, 125, and 250 counts per pixel.

This process allowed us to generate two kinds of completeness curves for
the GCs of each galaxy.  
Curves of the first kind represented the fraction of artificial clusters
that were successfully recovered by the task {\em daofind}.
Completeness curves of the second kind represented the fraction of artificial
clusters that were not only successfully recovered by {\em daofind}, but that
also returned a photometric magnitude within acceptable limits via the task
{\em phot}.  There were two of these limits: the returned photometric magnitude
was required to have an error of $<$0.3 mag and was also required to be
within 0.5 mag of the magnitude of the input artificial-cluster image.

A smoothed curve based on the completeness function given in \citet{fleming95}
was fit to each set of completeness curves generated.  The two
parameters returned by this fitting procedure were the 50\% completeness
limit, \kslim\ (expressed in magnitudes), and a steepness parameter $\alpha$.
Table~\ref{tablecplte} presents for each galaxy and background value the
two fit parameters derived, both for the first kind of completeness curves
(subscript ``fnd'' for ``found'') and for the second kind (subscript ``acc''
for ``accepted'').

\begin{figure}
  \begin{center}
    \includegraphics[width=0.4\textwidth]{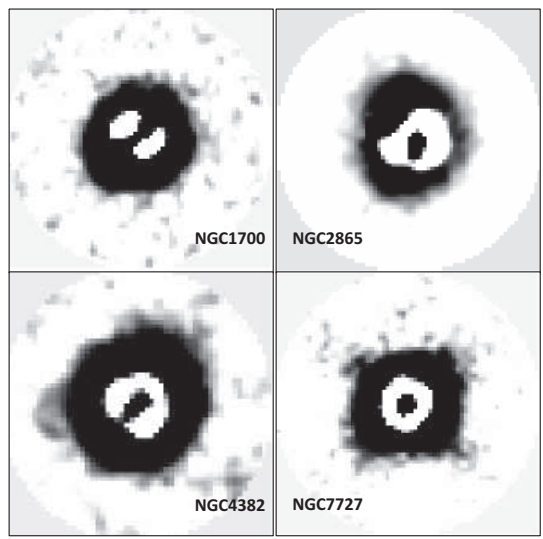}
    \caption{PSFs generated from the median background-subtracted images of
	     \n1700, \n2865, \n4382, and \n7727. The field of view is
	     $7.8\arcsec\times 7.8$\arcsec\ for each panel.}
    \label{fig:psf}
  \end{center}
\end{figure}

As mentioned above, several completeness curves were generated for each galaxy
in order to determine the stability of the photometric completeness limit
over a range of background values.  Given that the background varied
considerably over any one image, particularly across the center of the
galaxy, such stability was essential.  Indeed, the variance in background
over each image considerably affected the point-spread functions (PSFs) that
we attempted to construct.  In order to negate this effect, the final PSFs
that were used in the artificial cluster addition were generated from
background-subtracted GC images, as illustrated in Figure~\ref{fig:psf}.
These background-subtracted GC images were created using a $15\times 15$-pixel
median subtraction.

>From the results presented in Table~\ref{tablecplte} it is evident
that---although the steepness of the completeness curves did vary---the
50\% completeness
limit was exceptionally stable over the range of background values
investigated.  Therefore, we were able to use a single completeness limit
for each image, despite the fact that the background values varied
dramatically across the cores of the galaxies.

\subsubsection{\n1700}

\begin{figure}
  \begin{center}
    \includegraphics[angle=0,width=0.49\textwidth]{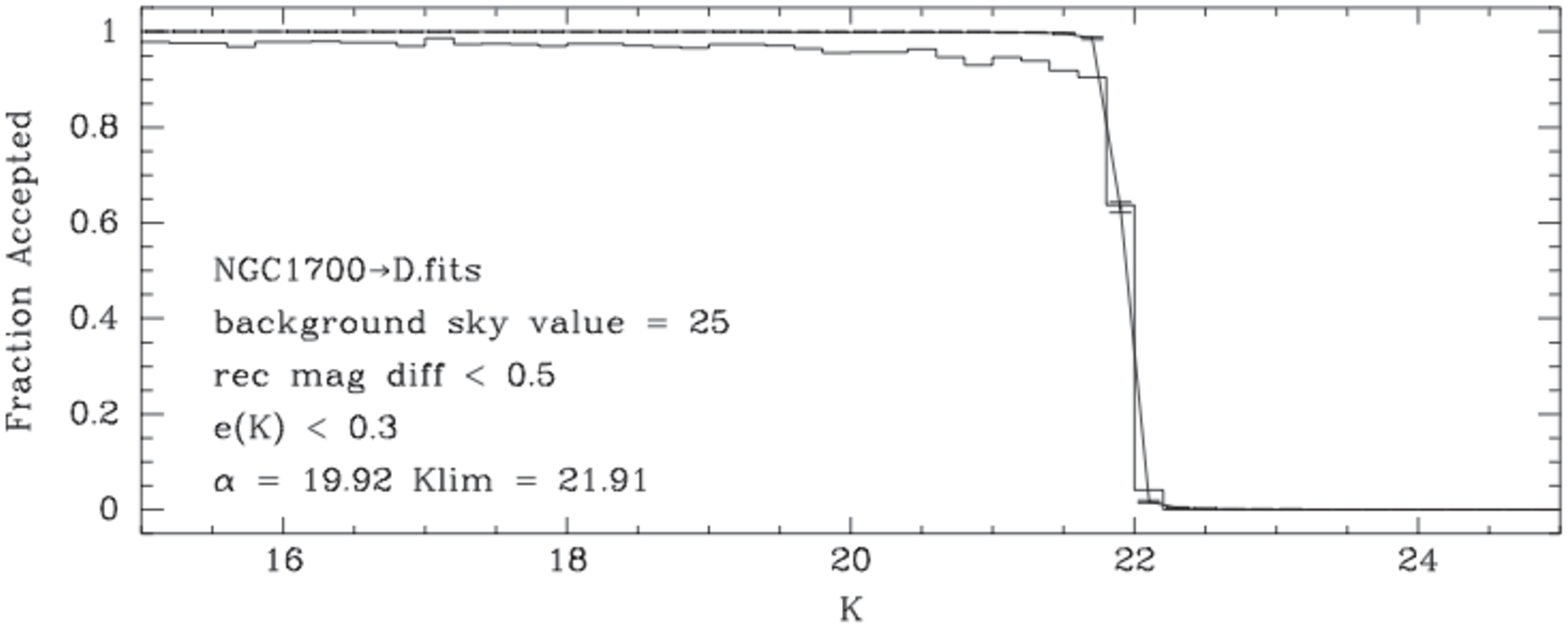}
    \caption{Completeness curve for \n1700 derived from artificial cluster
	     addition. The histogram shows the fraction of clusters recovered
	     to within acceptable error and magnitude limits, while the smooth
	     curve indicates the fitted completeness function.}
    \label{fig:NGC1700c}
  \end{center}
\end{figure}

Figure~\ref{fig:NGC1700c} shows the completeness curve derived for
GCs in \n1700 (shown for accepted artificial clusters only).
As Table~\ref{tablecplte} details, the 50\% \ks\ completeness limits were 
$\kslim = 21.90$\,--\,21.91 for both the recovered and the accepted clusters
at all background levels.
Hence, it is evident that photometric errors do not play a significant
role in setting the depth of completeness in this case.
Rather, the completeness is limited primarily by our ability to detect
the clusters.

\subsubsection{\n2865}

\begin{figure}
  \begin{center}
    \includegraphics[angle=0,width=0.49\textwidth]{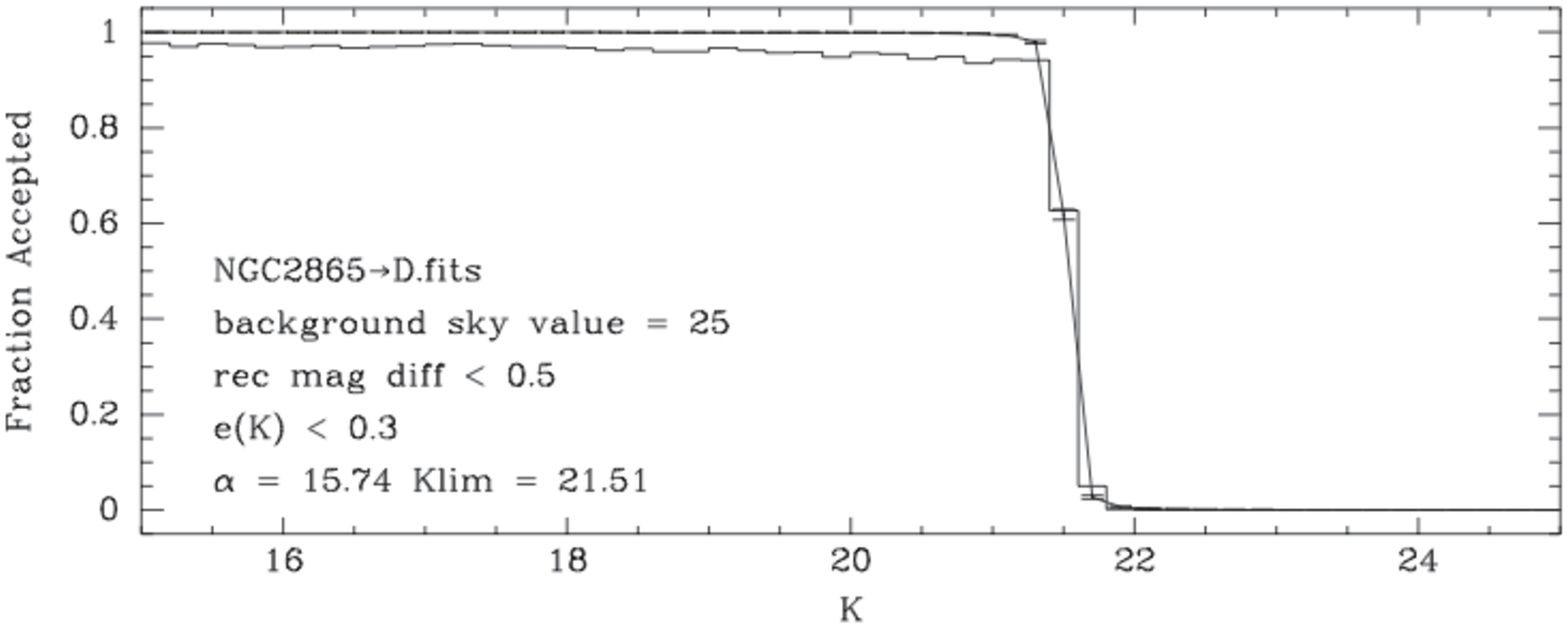}
    \caption{Completeness curve for \n2865 derived from artificial cluster
	     addition.  The histogram shows the fraction of clusters recovered
	     to within acceptable error and magnitude limits, while the smooth
	     curve indicates the fitted completeness function.}
    \label{fig:NGC2865c}
  \end{center}
\end{figure}

Figure~\ref{fig:NGC2865c} shows the completeness curve derived for GCs in
\n2865.
In this galaxy, the 50\% completeness limits were $\kslim = 21.51$\,--\,21.52
for both the recovered and the accepted artificial clusters.
As in the case of \n1700, the very small range of these values indicates that
the completeness in \ks\ was limited by our ability to detect the clusters
rather than by photometric errors.

\subsubsection{\n4382}

\begin{figure}
  \begin{center}
    \includegraphics[angle=0,width=0.49\textwidth]{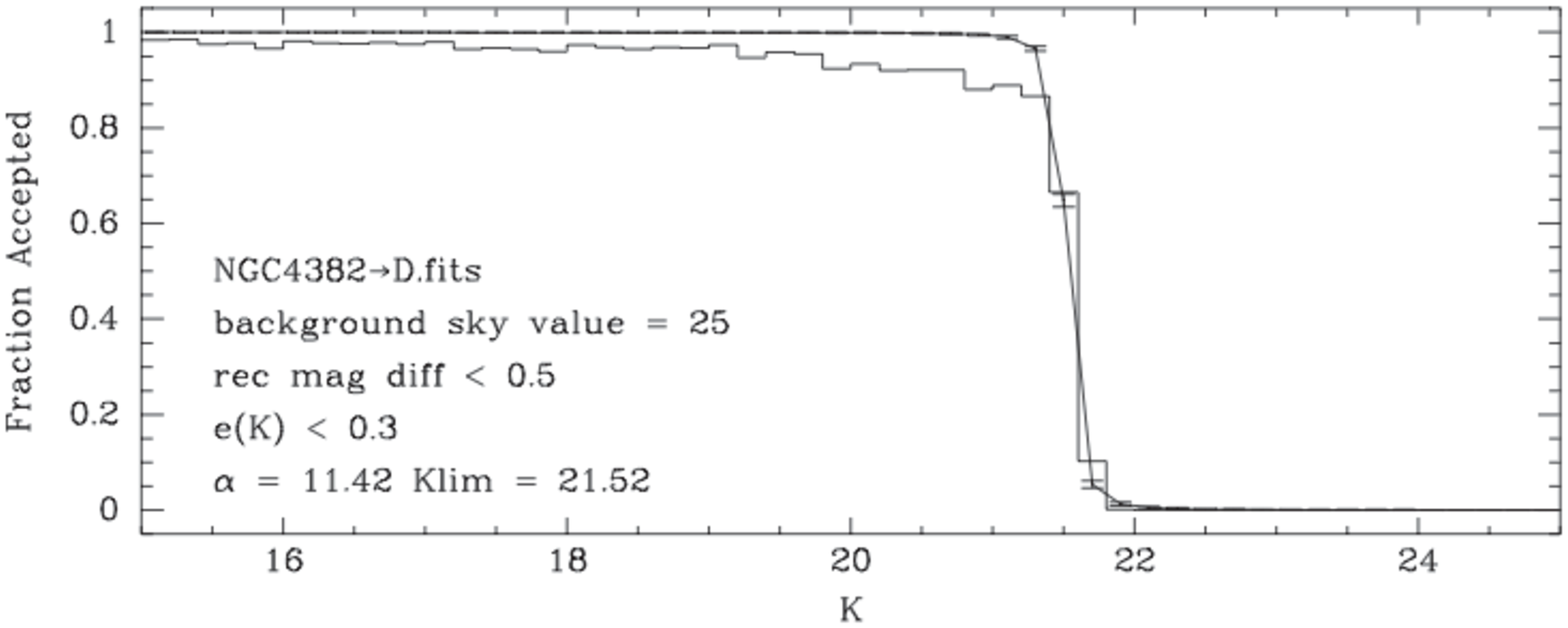}
    \caption{Completeness curve for \n4382 derived from artificial cluster
	     addition.  The histogram shows the fraction of clusters recovered
	     to within acceptable error and magnitude limits, while the smooth
	     curve indicates the fitted completeness function.}
    \label{fig:NGC4382c}
  \end{center}
\end{figure}

Figure~\ref{fig:NGC4382c} shows the completeness curve derived for GCs in
\n4382.
Again, the \ks\ 50\% completeness limits are similar for the recovered and
accepted curves, though with wider scatter than in \n1700 and \n2865 and
with a 0.06 mag difference between the two mean values (see
Table~\ref{tablecplte}).

\subsubsection{\n7727}

\begin{figure}
  \begin{center}
    \includegraphics[angle=0,width=0.49\textwidth]{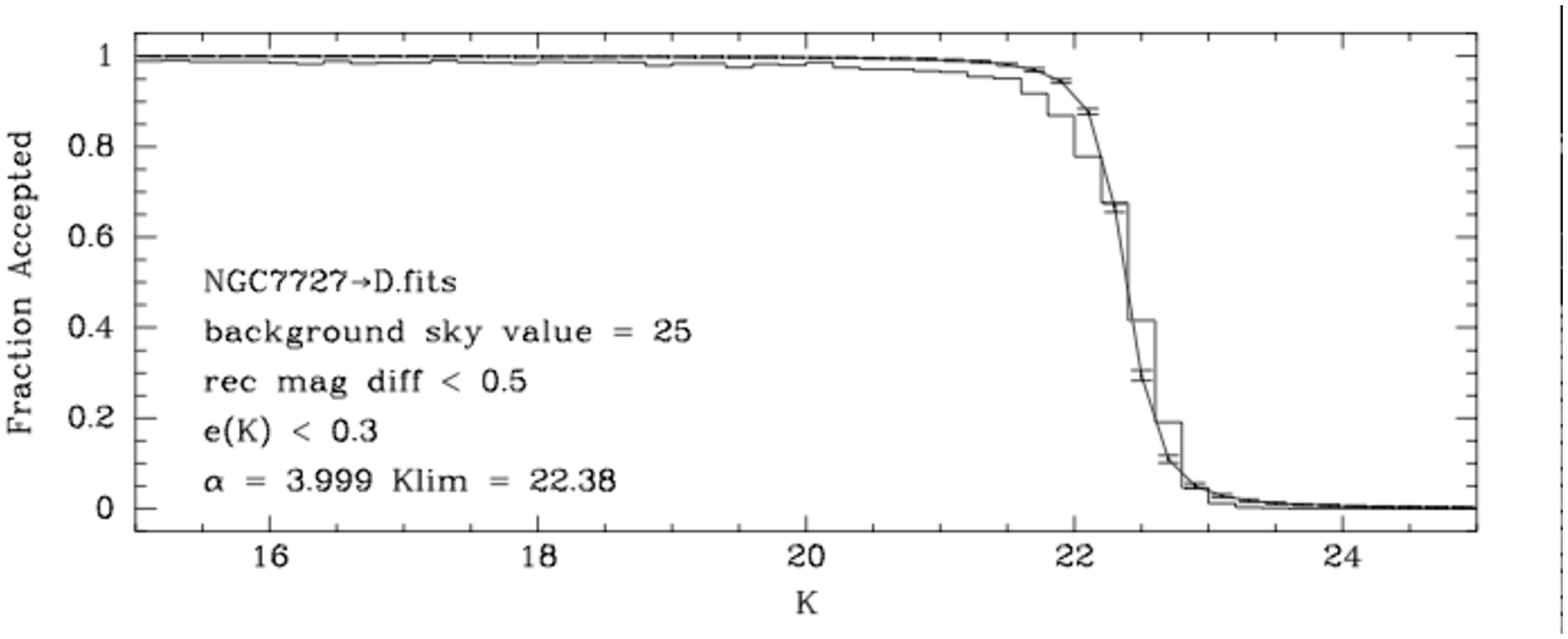}
     \caption{Completeness curve for \n7727 derived from artificial cluster
	     addition.  The histogram shows the fraction of clusters recovered
	     to within acceptable error and magnitude limits, while the smooth
	     curve indicates the fitted completeness function.}
    \label{fig:NGC7727c}
  \end{center}
\end{figure}

Figure~\ref{fig:NGC7727c} shows the completeness curve derived for GCs in
\n7727.

As Table~\ref{tablecplte} details, the various \ks\ 50\% completeness limits
scatter more and show a significant systematic difference of 0.09 mag between
the recovered artificial clusters (mean $\kslimfnd = 22.46$) and the accepted
clusters (mean $\kslimacc = 22.37$).
Notice that these limits for the \n7727 GCs are more than 0.5 mag fainter
than the limits for GCs in the other three galaxies, which is not surprising
given that the \ks\ images of \n7727 were obtained with FLAMINGOS-I on a night
of particularly good seeing.
It is because of these fainter limits that the photometric errors are more
influential in \n7727 than in the other three galaxies.

Nevertheless, we adopted the single value of $\kslimacc = 22.37$ as our
acceptance criterion for candidate GCs in \n7727.

%%%%%%%%%%%%%%%%%%%%%%%  Table_2 %%%%%%%%%%%%%%%%%%%%%%%%

\begin{deluxetable}{lccccc}
\def\psn{\phs\phn}
\def\pnn{\phn\phn}
\tablecaption{Summary of the parameters for the completeness curves for each
	      galaxy.}
\tablehead{
  \colhead{Galaxy}                                  &
  \colhead{Sky Value}                               &
  \colhead{\kslimfnd\tablenotemark{a}}              &
  \colhead{$\alpha_{\rm fnd}$\tablenotemark{b}}     &
  \colhead{\kslimacc\tablenotemark{a}}              &
  \colhead{$\alpha_{\rm acc}$\tablenotemark{b}}     \\
  \colhead{NGC\ \ \ } &
  \colhead{(counts)}  &
  \colhead{(mag)}     &
  \colhead{}          &
  \colhead{(mag)}     &
  \colhead{}
}
\startdata
 1700 &  \phn25 &  21.91 &    22.05 & 21.90 &    19.27 \\
 1700 &  \phn55 &  21.91 &    15.74 & 21.90 &    12.84 \\
 1700 &  \phn85 &  21.91 &    21.19 & 21.90 & \phn9.17 \\
 1700 &     125 &  21.91 &    18.41 & 21.90 &    18.97 \\
 1700 &     250 &  21.91 &    17.62 & 21.90 &    16.33 \\
 2865 &  \phn25 &  21.52 &    16.00 & 21.51 &    15.10 \\
 2865 &  \phn55 &  21.52 &    11.81 & 21.51 &    12.81 \\
 2865 &  \phn85 &  21.51 &    16.67 & 21.51 &    13.33 \\
 2865 &     125 &  21.52 &    15.39 & 21.51 & \phn7.09 \\
 2865 &     250 &  21.52 &    14.63 & 21.51 & \phn8.37 \\
 4382 &  \phn25 &  21.55 &    11.57 & 21.51 &    10.55 \\
 4382 &  \phn55 &  21.56 &    11.18 & 21.46 & \phn3.68 \\
 4382 &  \phn85 &  21.55 &    11.87 & 21.51 &    14.11 \\
 4382 &     125 &  21.57 &    10.49 & 21.51 &    10.03 \\
 4382 &     250 &  21.54 &    13.03 & 21.47 & \phn4.01 \\
 7727 &  \phn25 &  22.46 & \phn3.93 & 22.38 & \phn3.99 \\ 
 7727 &  \phn55 &  22.45 & \phn3.76 & 22.37 & \phn4.44 \\
 7727 &  \phn85 &  22.46 & \phn4.26 & 22.38 & \phn3.81 \\
 7727 &     125 &  22.46 & \phn3.94 & 22.37 & \phn3.55 \\
 7727 &     250 &  22.45 & \phn3.96 & 22.35 & \phn3.55
\enddata

\tablenotetext{a}{50\% \ks\ completeness limit.}
\tablenotetext{b}{Curve fitting parameter \citep{fleming95} governing
		  the slope of the fit.}
\label{tablecplte}
\end{deluxetable}

\subsection{Further Corrections}

\begin{figure*}
  \begin{center}
    \includegraphics[angle=0,width=1\textwidth]{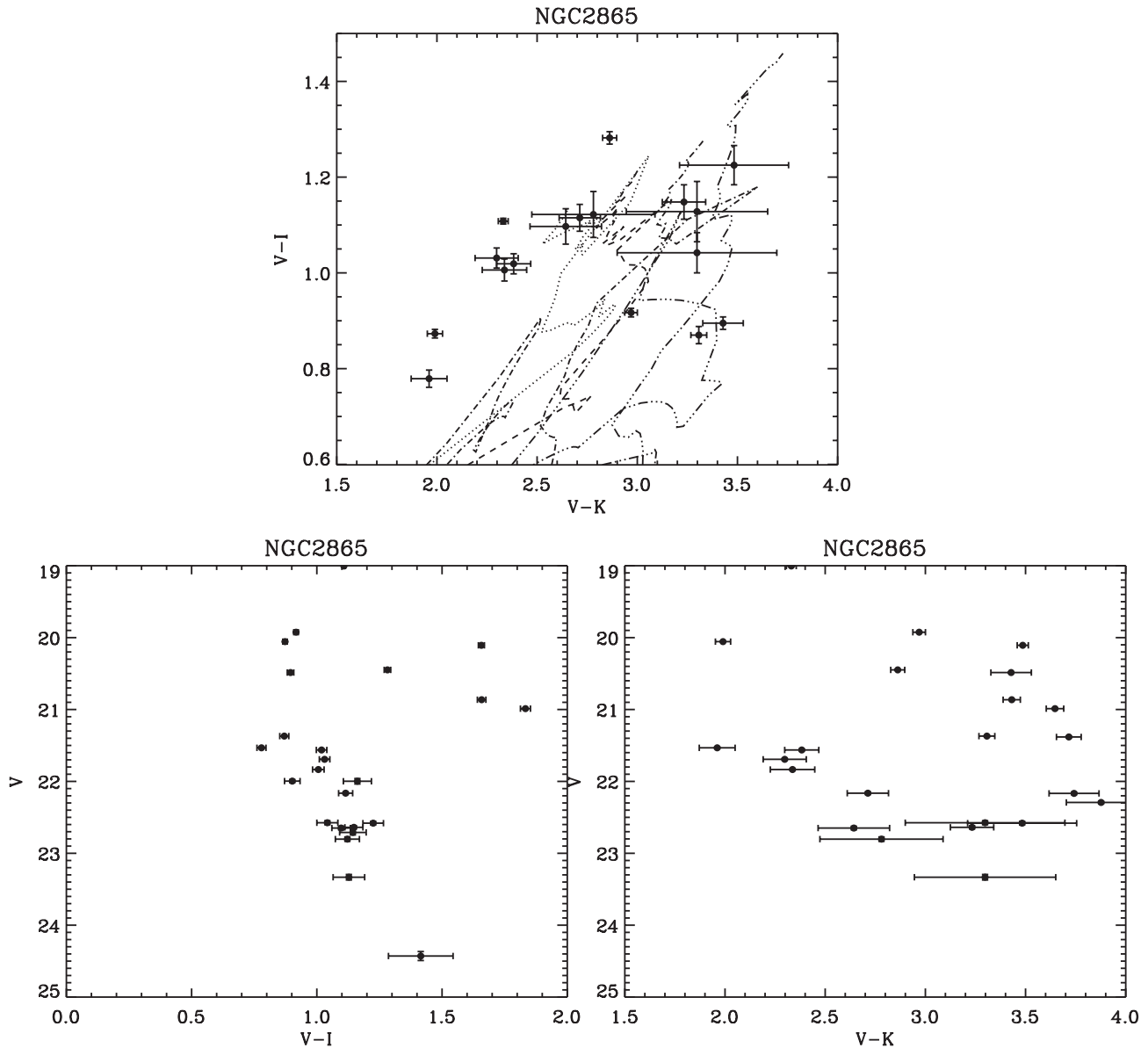}
    \caption{Color--magnitude and color--color diagrams for GCs of \n2865.}
    \label{fig:NGC2865VIK}
  \end{center}
\end{figure*}

\begin{figure*}
  \begin{center}
    \includegraphics[angle=0,width=1\textwidth]{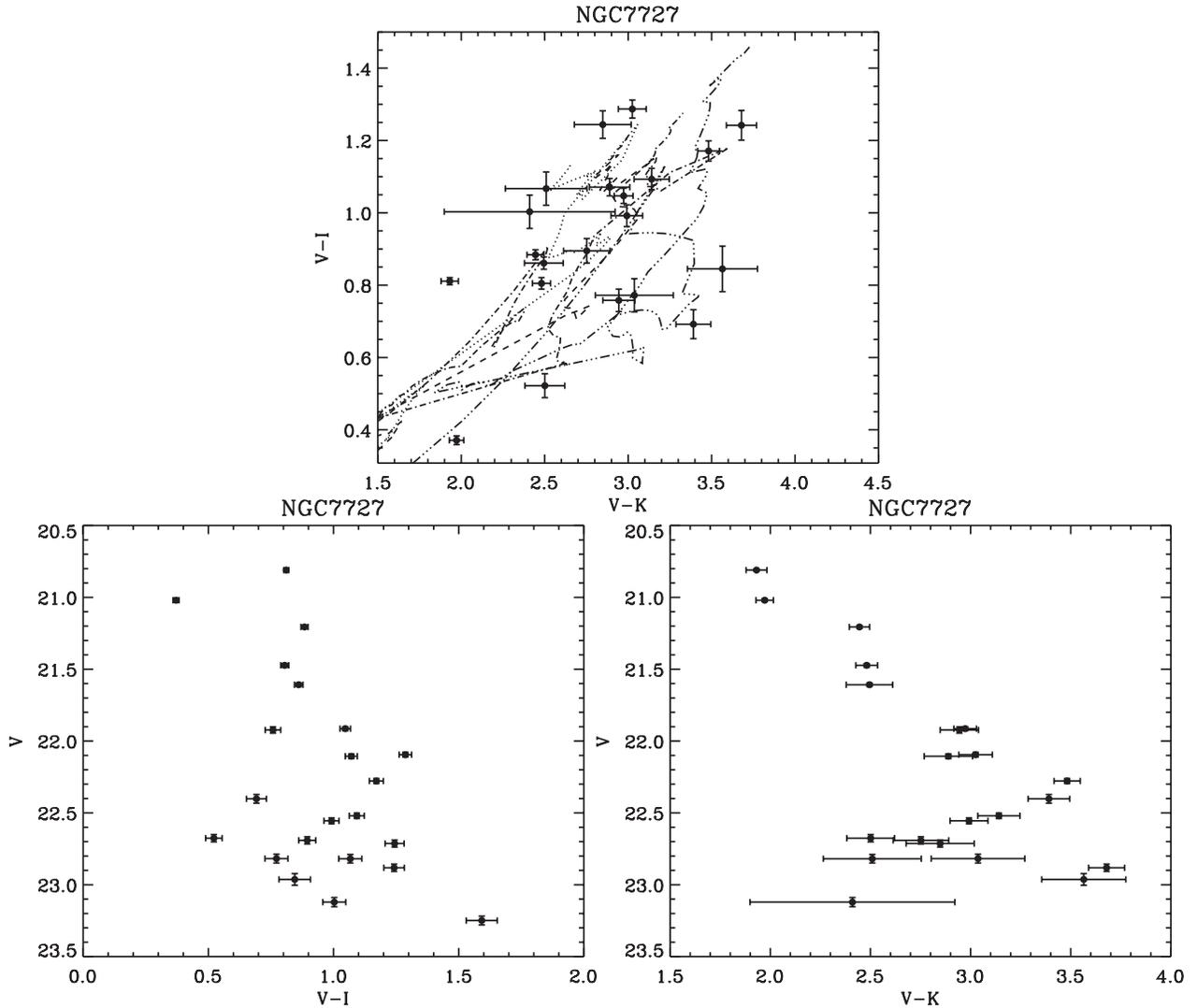}
    \caption{Color--magnitude and color--color diagrams for GCs of \n7727.}
    \label{fig:NGC7727VIK}
  \end{center}
\end{figure*}

In addition to the completeness corrections we performed a visual examination
of each GC candidate, using the IRAF task {\em imexam} to further remove
any contaminants such as background galaxies.  Candidates were dropped if
their { shape} was particularly elliptical and/or if the counts per pixel
within the object were well below those of the other candidates.
Table~\ref{table:candidates} gives a summary of the GC candidate corrections
applied for each GC system.
Figures~\ref{fig:NGC2865VIK} and \ref{fig:NGC7727VIK} show the resulting
color--color and color--magnitude diagrams for the GC systems of \n2865
and \n7727, respectively.

%%%%%%%%%%%%%%%%%%%%%%%  Table_3 %%%%%%%%%%%%%%%%%%%%%%%%

\begin{deluxetable}{lcccc}
\tablecaption{Summary of globular-cluster candidates removed during
	each correction}
\tablehead{
  \colhead{}             &  
  \colhead{Initial}      &  
  \colhead{Removed}      &  
  \colhead{Removed}      &  
  \colhead{Final}           \\
  \colhead{Galaxy}       &
  \colhead{GC}      &
  \colhead{via Com-}     &
  \colhead{via Visual}   &
  \colhead{GC}              \\
  \colhead{NGC\ \ \ }    &
  \colhead{Candidates}   &
  \colhead{pleteness}    &
  \colhead{Inspection}   &
  \colhead{Candidates}
}
\startdata
1700 &  45 &  6 &    13 &  26 \\
2865 &  65 &  2 &    20 &  43 \\
4382 &  65 &  3 & \phn2 &  60 \\
7727 &  31 &  0 & \phn9 &  22
\enddata
\label{table:candidates}
\end{deluxetable}

\section{FITTING TO SSP MODELS}\label{sec:results}

In order to constrain the age of the merger event for each of the four
galaxies, the ages of their GC subpopulations were determined.
Specifically, we assumed that the parameters of the youngest subpopulation
may be associated with the most recent major star-formation episode---the
galaxy-merger event itself.

Conventional studies of GC ages and metallicities are often based on
optical-to-NIR color--magnitude diagrams such as, e.g., the $V$ vs \vi\
diagram.
However, such diagrams suffer from an inherent age--metallicity degeneracy
\citep[e.g.,][]{worthey99}, and it is often difficult to disentangle age and
metallicity without additional detailed spectra.

In an effort to break---or at least partly lift---this degeneracy, we
use \viks\ color--color diagrams (specifically \vi\ vs \vks) and interpret
the measured GC colors via the SSP models generated by Charlot \& Bruzual
(2007, private communication, hereafter CB07; see \citealt{bc03} and
\citealt{bruz07} for published models and details).

The fitting algorithm used to match the models to our data is briefly
described below and closely follows the method and algorithm used by
\citet{hempel04}.
We also present the results of our efforts to verify the technique.

\subsection{Methodology}

Our principal goal was to identify the possible presence of multiple
subpopulations among the GCs of our sample galaxies and to estimate
their ages, based on simple assumptions about their metallicities.

We assume that there are three predominant cluster subpopulations of
interest: one consisting of $\sim$13 Gyr old metal-poor GCs, a second
one of also old ($>$10 Gyr) metal-rich GCs, and a third one of
younger, intermediate-age metal-rich GCs.
The universal old metal-poor subpopulation is present in all major
galaxies and is represented in our model simulations by clusters
of mean logarithmic metallicity $[Z/Z_{\odot}] = -1.7$, with \viks\
colors computed by CB07; the size and age of this subpopulation is being
held constant during the simulations for each galaxy.
The two metal-rich subpopulations are assumed to both have logarithmic
metallicities in the range of $-0.7 < [Z/Z_{\odot}] < 0.4$.
The number ratios of metal-rich to metal-poor clusters, and of
intermediate-age to old clusters within the metal-rich subpopulation, are
then varied systematically.

With these assumptions made, the goal is to simulate the appearance of
the mixed GC populations in \vi\ vs \vks\ diagrams with BC07 model clusters
and to then compare these model diagrams with the observed diagrams for
the GCs of each galaxy.
The measure used for the comparison is provided by the cumulative age
distributions of the artificial clusters and the observed GCs, both
derived from the positions of the clusters in their respective \vi\ vs
\vks\ diagrams and compared via a chi-squared (hereafter \chisq) test
described below.

To {\em simulate\,} the observed mixed GC population of each galaxy, a set
of model cluster \vks\ colors was generated using a Monte Carlo technique.
For each of the many simulations per galaxy, the total number of model
clusters was made to equal the number of GC candidates in the galaxy
(Table~\ref{table:candidates}, last column).
Corresponding model cluster \vi\ colors were then computed based on the
SSP models of CB07.
To simulate observational errors, each model cluster population was
furthermore convolved with photometric errors based on the indicative errors
of our data.
Finally, cumulative age distributions were calculated for each model cluster
population, based on the same SSP models by CB07.

For each of our four galaxies, we generated a grid of 121 model cluster
populations as follows.
The age of the old metal-rich subpopulation was held fixed at 13 Gyr, while
the age of the younger metal-rich subpopulation was set to 0.2, 0.5, 0.75, 1,
1.5, 2, 3, 5, 7, 10, and 13 Gyr.
For each of these 11 ages, the ratio of young to old clusters was varied from
0\% to 100\% in 10\% steps, thus yielding a total of 121 model populations
for comparison with the observed GC population of each galaxy.

For each of the 121 age and number-ratio combinations, 1000 model samples
were generated via the Monte Carlo technique, and the average
{ cumulative}
age distribution (based on the SSP models) of the simulated clusters was
calculated.  The
{ cumulative}
age distribution of the observed GCs was also calculated based on the same
SSP models, via a transformation from the \vi\ vs \vks\ plane to the
abundance vs log-age plane based on these models (for details, see
\citealt{hempel04}).

We then performed \chisq\ fitting { of the model cumulative age
distributions to the observed cumulative age distribution} to determine
which of the age and number-ratio combinations best fit the data.
For each galaxy, a contour plot of the \chisq\ values over the age vs.\
number-ratio configuration plane was drawn and used to infer the age and
number ratio of the young metal-rich subpopulation as compared to the old
subpopulation.
These contour plots are shown and discussed in \S~\ref{sec43} below.

\begin{figure}
  \begin{center}
    \subfigure{\includegraphics[width=0.4\textwidth]{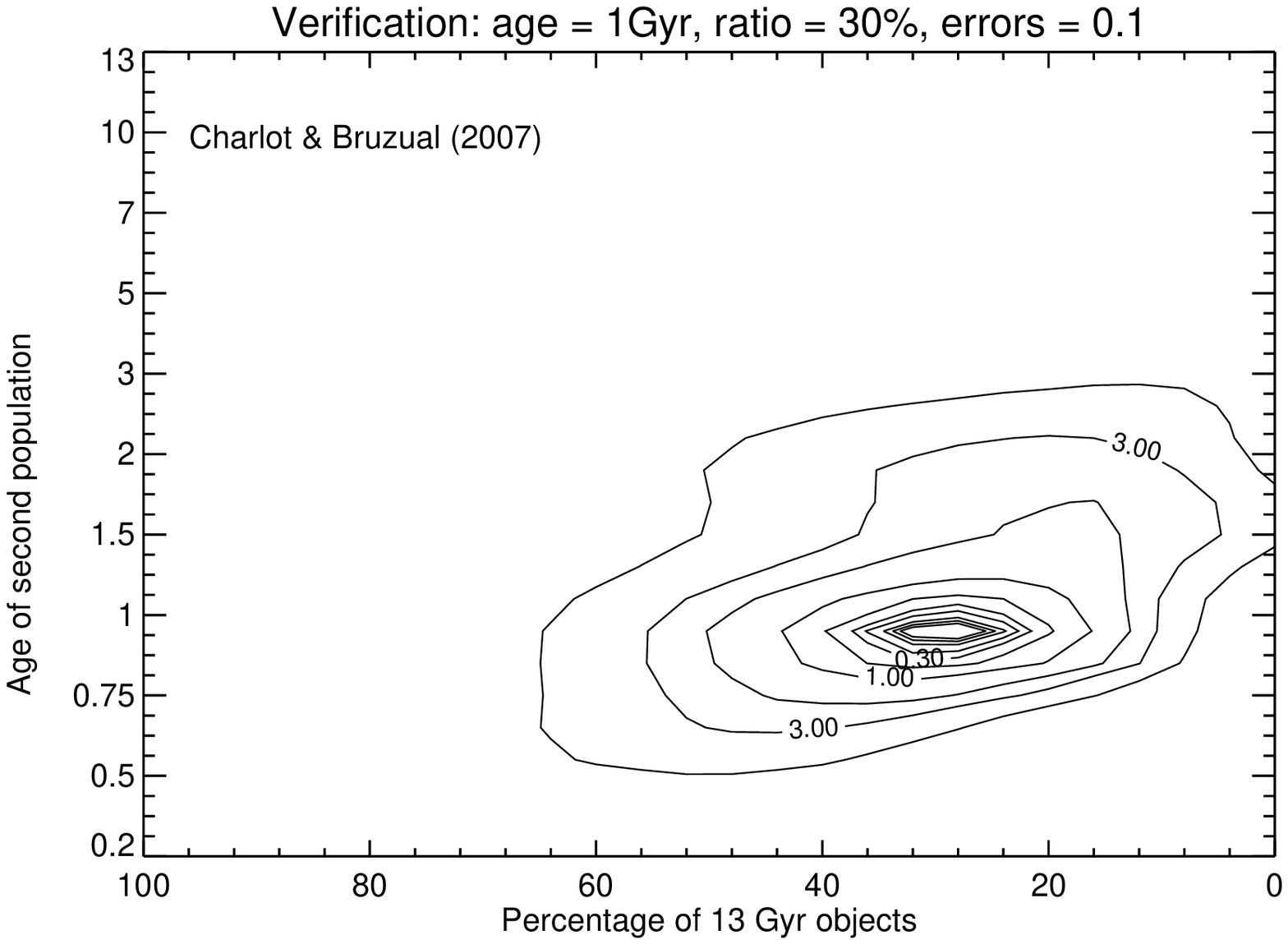}}
    \subfigure{\includegraphics[width=0.4\textwidth]{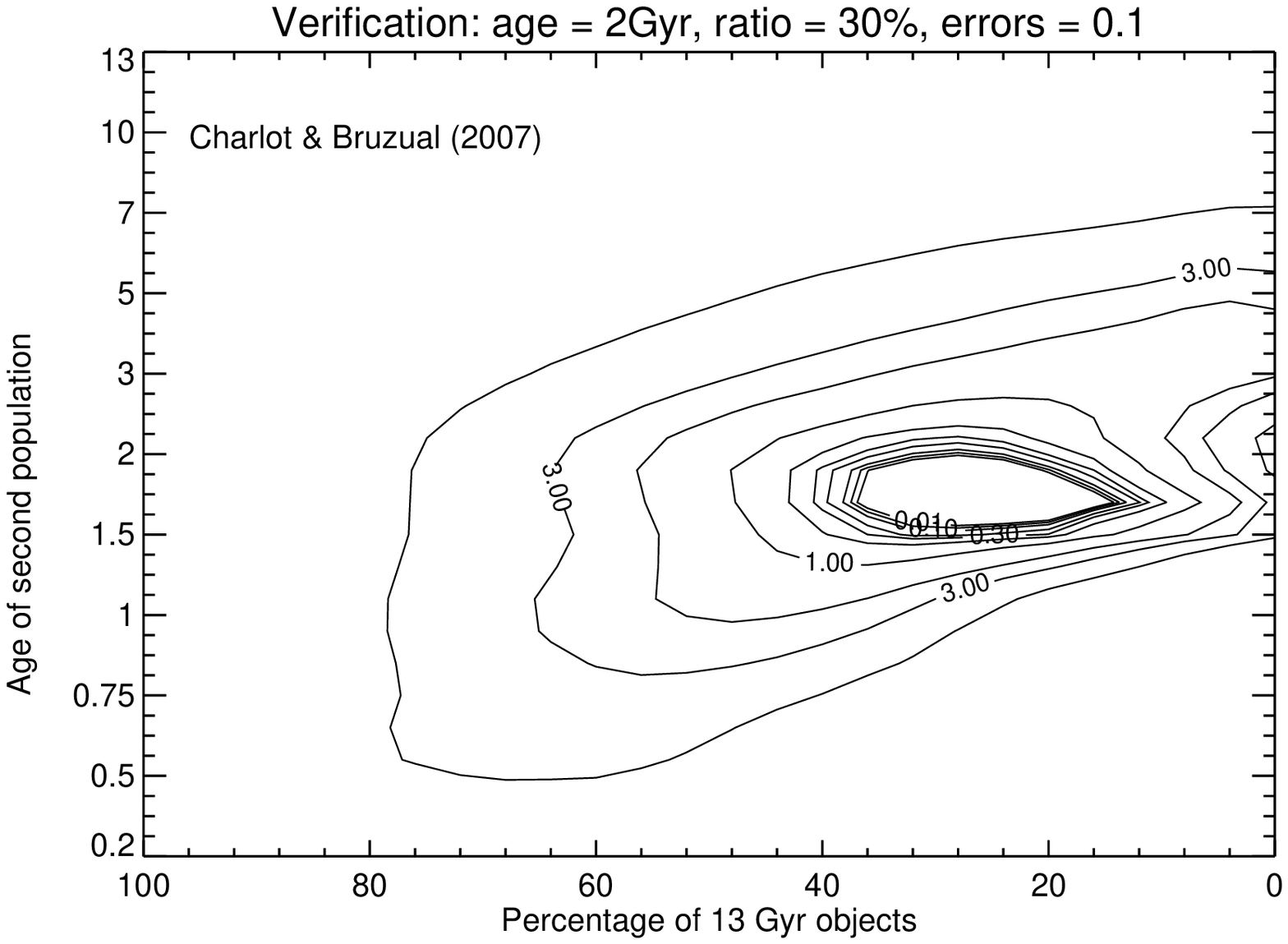}}
    \subfigure{\includegraphics[width=0.375\textwidth]{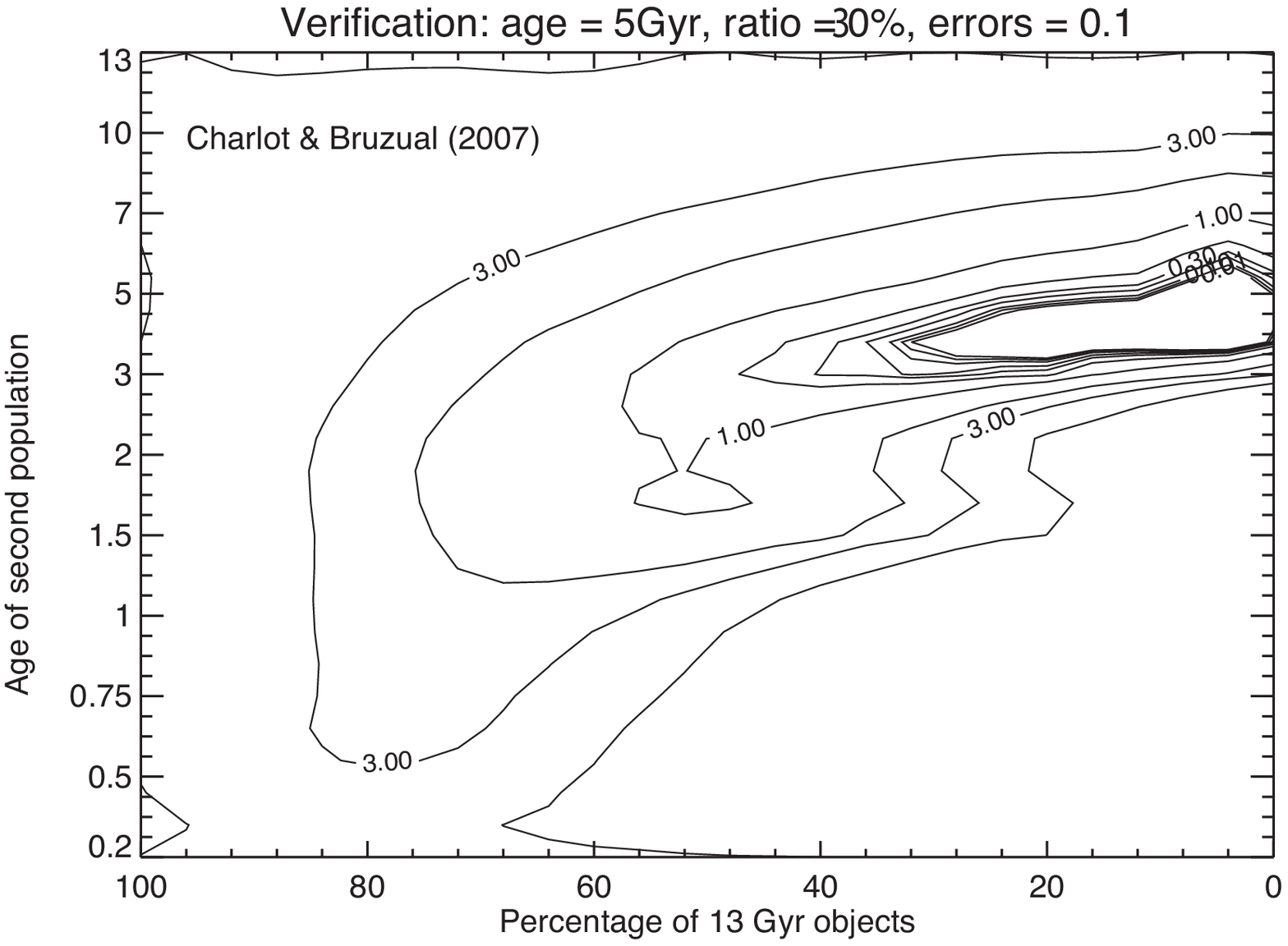}}
    \caption{\chisq\ contour plots using a simulated cluster population as
	     input.  The age of the younger subpopulation increases from
	     1 Gyr (top) to 2 Gyr (middle) and 5 Gyr (bottom).  The input age,
	     number ratio, and photometric error are given for each.}
    \label{fig:chi2_trial}
  \end{center}
\end{figure}

\begin{figure}
  \begin{center}
    \subfigure{\includegraphics[width=0.4\textwidth]{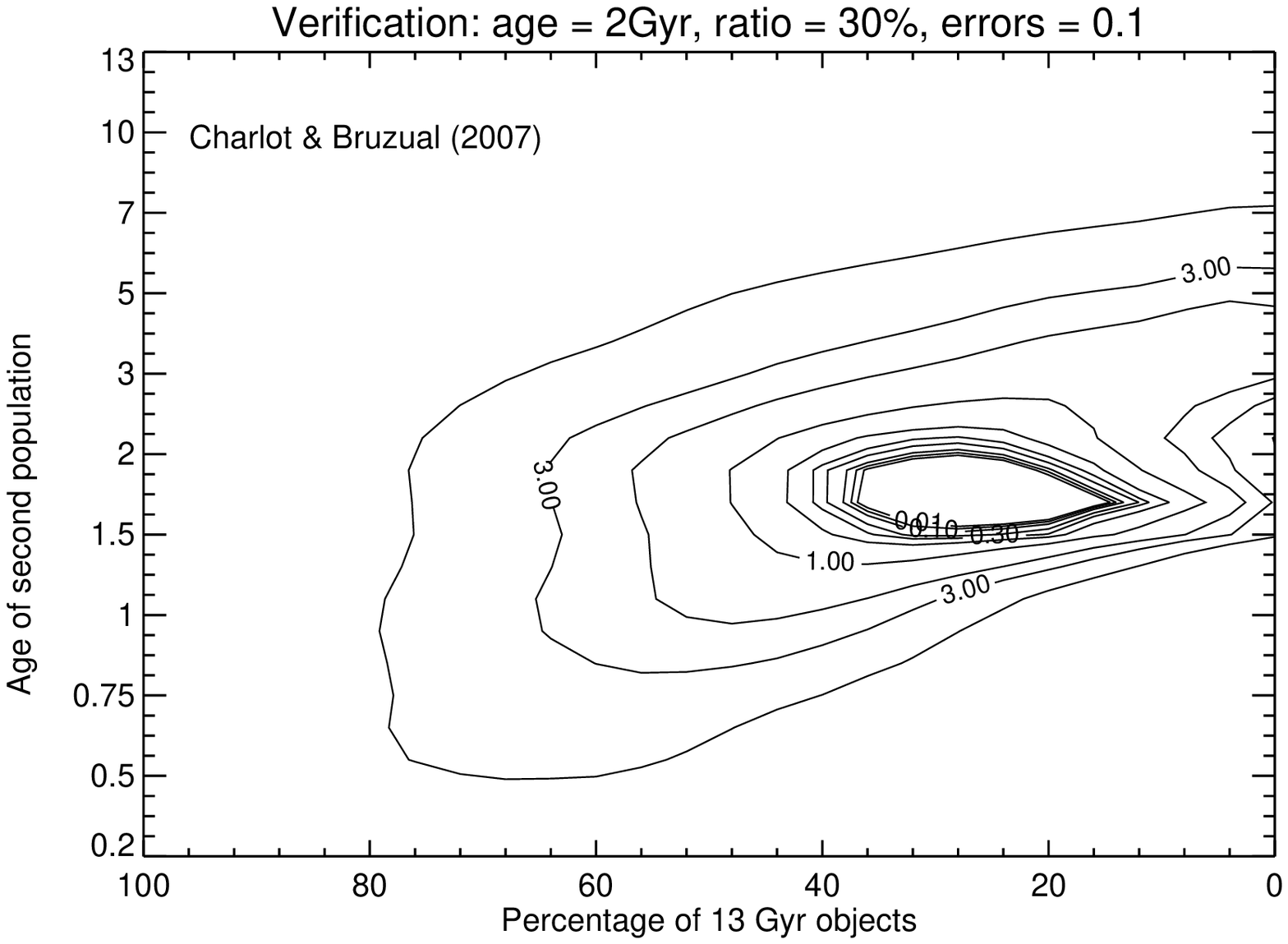}}
    \subfigure{\includegraphics[width=0.4\textwidth]{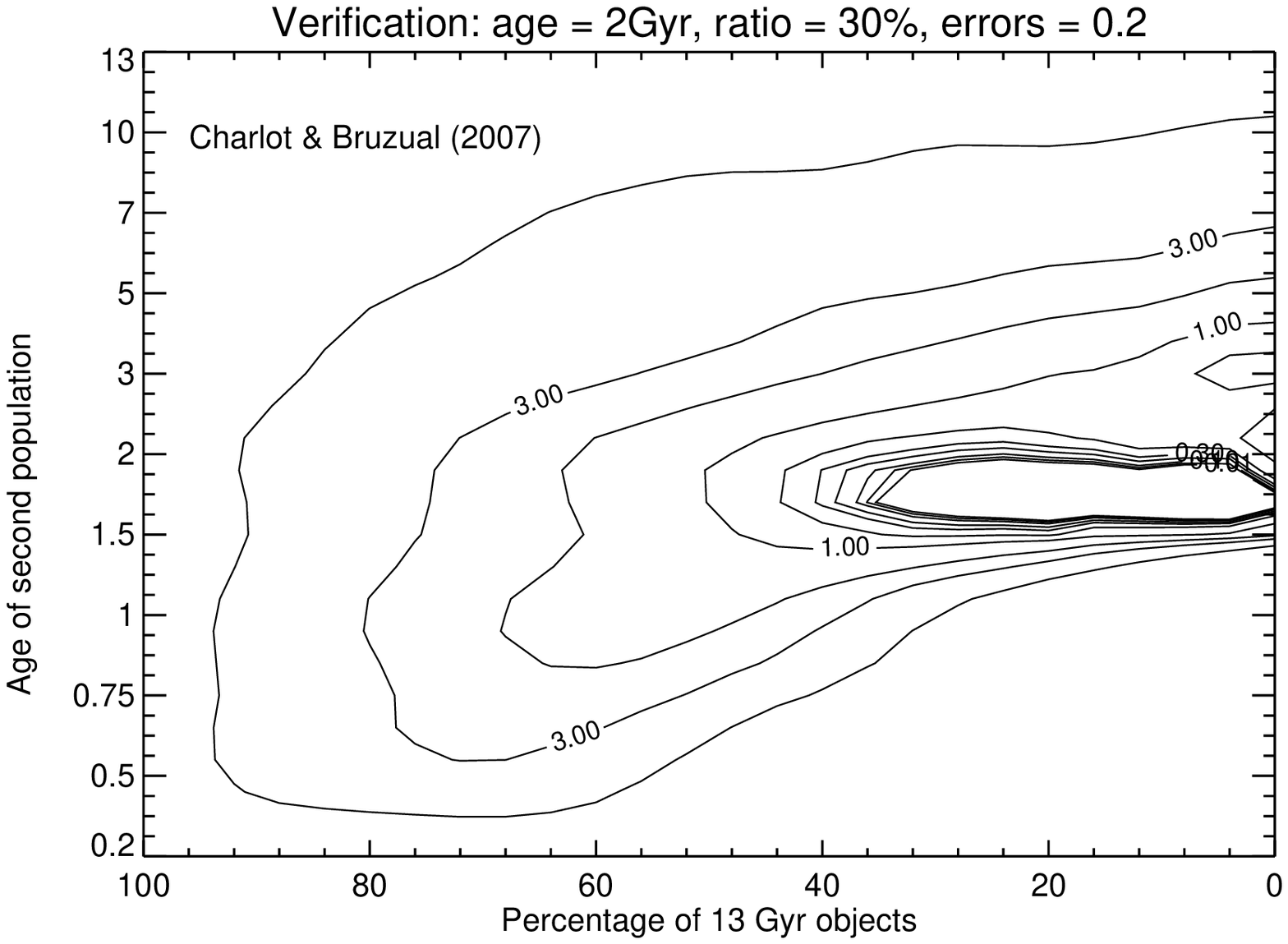}}
    \subfigure{\includegraphics[width=0.4\textwidth]{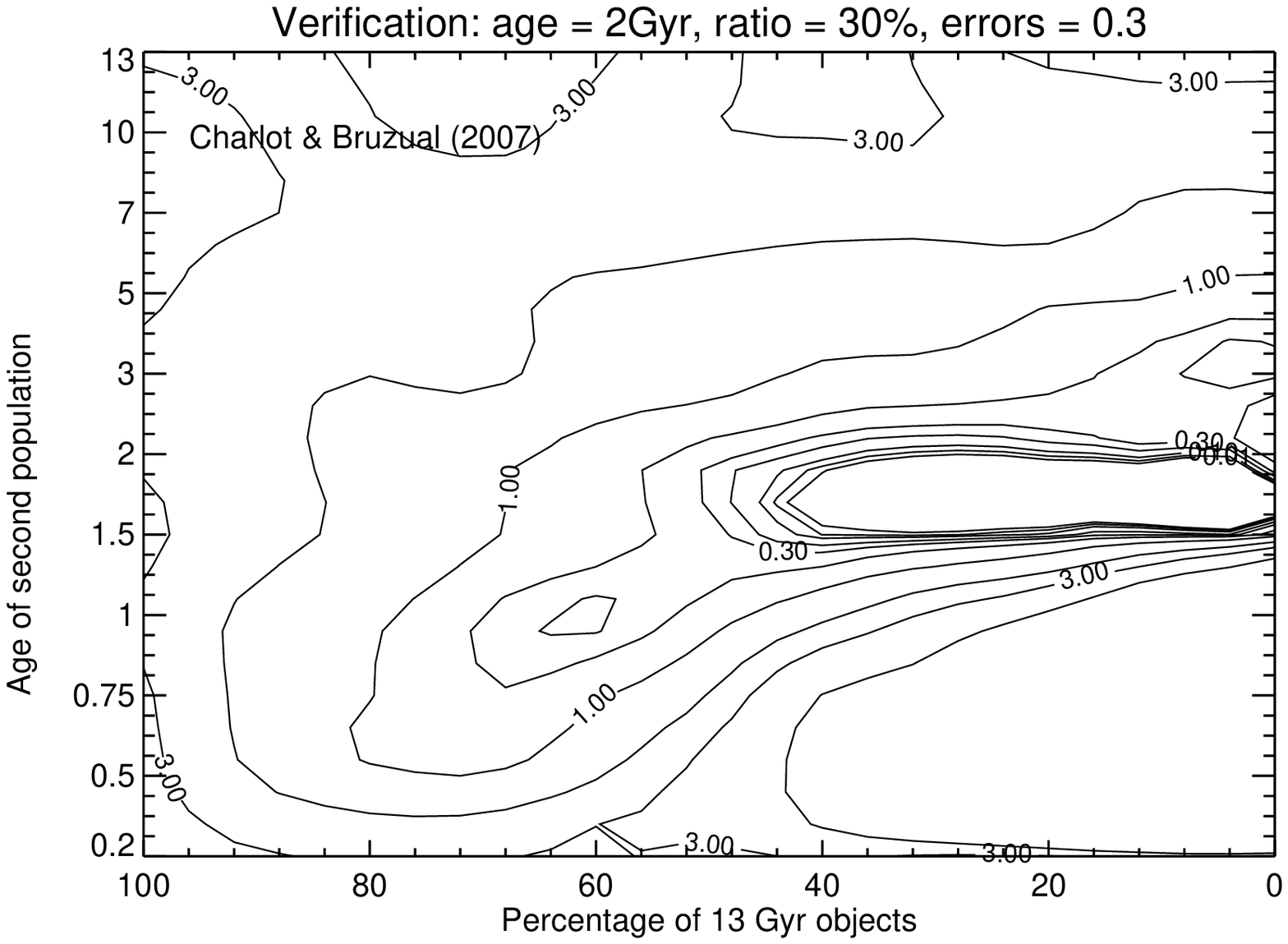}}
    \caption{\chisq\ contour plots using a simulated cluster population as
	     input.  The size of the photometric errors increases from 0.1 mag
	     (top) to 0.2 mag (middle) and 0.3 mag (bottom).  The input age,
	     number ratio, and photometric error are given for each.}
    \label{fig:chi2_trial2}
  \end{center}
\end{figure}

\subsection{Verification of Technique}

In order to validate the use of this technique, we performed a series of
verification tests independent from those performed by \citet{hempel04}.
Our verification method involved replacing the observed GC population
with a simulated population of model clusters.
The simulation of this population followed the same procedure as the
simulations of our model populations described above.

First, we tested the ability of our adopted method to return an accurate
value for the age of the younger metal-rich input population.
As Figure~\ref{fig:chi2_trial} shows, while the age returned by the
\chisq-fitting did not converge exactly to the simulated input age, it was
largely consistent (to within $\sim$10\%).
Hence, we can confirm that, over the range of ages ($\sim$\,1\,--\,5 Gyr)
that we are investigating, the method returns a reasonably accurate age for
a given input age.

Second, we also tested our method's stability with respect to the photometric
errors of the input data.  For a given age and ratio configuration, we varied
the photometric errors from 0.1 mag to 0.3 mag.  Note that (i) the indicative
photometric errors in our observed data were below 0.3 mag, and (ii) while
there were some clusters with higher errors, they were cut before the
\chisq-fitting was performed.

Figure~\ref{fig:chi2_trial2} highlights that, as expected, increasing the
errors decreased the convergence of the \chisq\ fit.  Even so, however, the
age returned by the fitting contours was still largely consistent with the
simulated input age.  Thus, we confirm the validity of our method for both
the ages and the indicative photometric errors of the input data.

\begin{figure*}
  \begin{center}
    \subfigure{\includegraphics[width=0.85\textwidth]{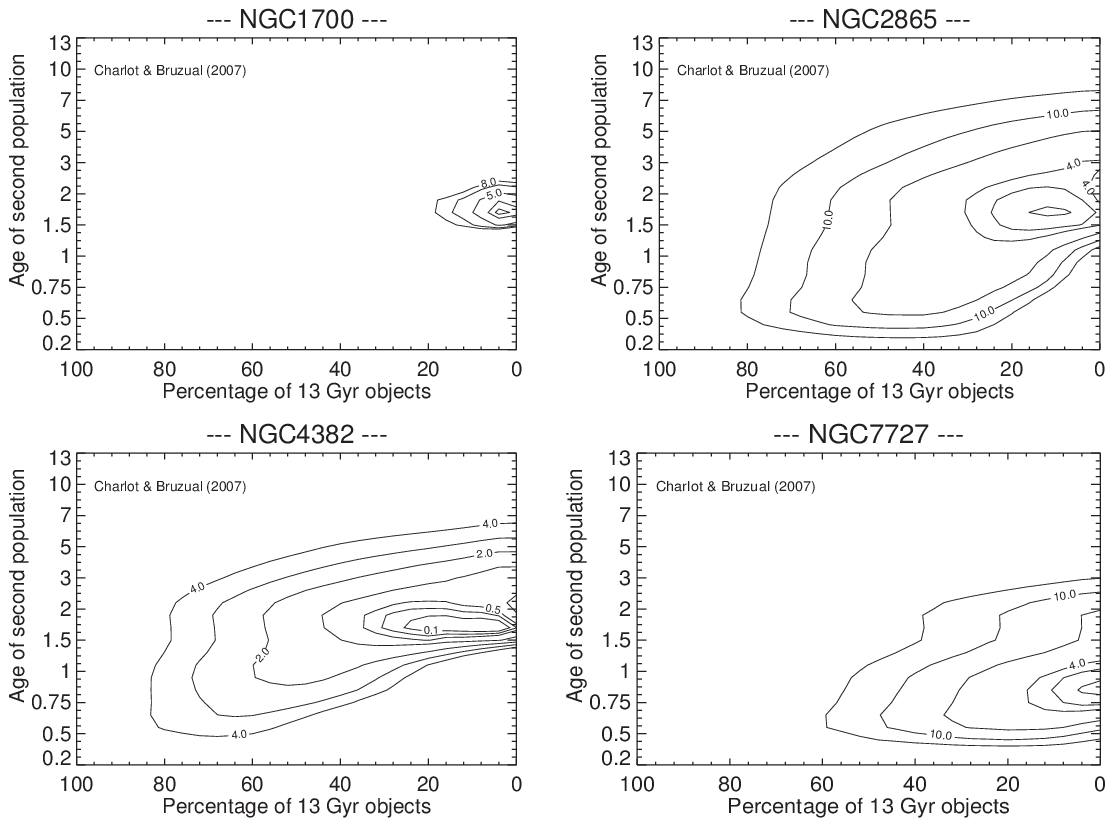}}
    \caption{Contour plots of \chisq\ values from fits of SSP model colors
	     (CB07) to our \viks\ colors for GCs in \n1700, 2865, 4382, and
	     7727. 
	     Notice that the percentages of 13 Gyr old GCs among those
	     observed are small.  Most observed GCs in the four sample
	     galaxies belong to younger, metal-rich subpopulations of
	     intermediate age (see Table~\ref{table4}).}
    \label{fig:chi2_contour}
  \end{center}
\end{figure*}

>From the spread in the \chisq\ contours and the deviation of the converged
age from the input age we infer an upper limit for the age uncertainty of
about $\pm 0.75$ Gyr.  Similarly, for the young-to-old ratio we find an 
upper limit to the uncertainty of about $\pm 20\%$.  These uncertainties
will be used for the GC subpopulations of each galaxy in the following
sections.

\subsection{Cluster Ages}
\label{sec43}

Figure~\ref{fig:chi2_contour} shows contour plots of \chisq\ values over the
modeled configuration space for each GC system of our four sample galaxies.
In each contour plot, the convergence of \chisq\ to a minimum value marks the
most likely combination of age and number-ratio for the younger GC
subpopulation of that galaxy, based on the SSP models by CB07.
Table~\ref{table4} summarizes these resulting combinations, which we now
briefly discuss for each galaxy.

\subsubsection{\n1700}

For this galaxy, Figure~\ref{fig:chi2_contour} shows that the \chisq\ values
converge to a minimum at a GC population mix consisting of about 5\% of
an old subpopulation and 95\% of a younger subpopulation with an age of
$1.7 \pm 0.8$ Gyr.

Note that this age is consistent with the previously published age estimate
of $3\pm 1$ Gyr \citep{brown00} to within the combined errors.
Note also that the quoted percentages apply only to the {\em observed}
GC candidates, and not to the entire GC population, whose total number and
subpopulation percentages are unknown.

\subsubsection{\n2865}

For \n2865, the \chisq\ contours again show a high level of convergence, as
can be seen in the top-right panel of Figure~\ref{fig:chi2_contour}.
From the contour minimum we infer a younger metal-rich GC subpopulation of
age $1.8 \pm 0.8$ Gyr, comprising about 90\% of the observed GC candidates.

Again, this newly determined age for the younger GCs compares well with the
previously published age of $1.8\pm 0.5$ Gyr for the young {\em stellar\,}
subpopulation in this galaxy \citep{rampazzo07}.

%%%%%%%%%%%%%%%%%%%%%%%  Table_4  %%%%%%%%%%%%%%%%%%%%%%%%

\begin{deluxetable*}{lcccl}
\def\psn{\phs\phn}
\def\pnn{\phn\phn}
\tablewidth{0pt}
\tabletypesize{\scriptsize}
\tablecaption{Results of \chisq\ fitting: Ages and percentages \\
	      of intermediate-age GC subpopulations}
\tablehead{
  \colhead{Galaxy}        &
  \colhead{Age\tablenotemark{a}} &
  \colhead{Percentage}    &
  \colhead{Published Age} &
  \colhead{Source}          \\
  \colhead{NGC\ \ \ }     &
  \colhead{(Gyr)}         &
  \colhead{Young\tablenotemark{b}} &
  \colhead{(Gyr)}         &
  \colhead{}
}
\startdata
1700 &  1.7$\pm$0.8 &    95\% &  $3\pm 1.0$   &  \citet{brown00}    \\
2865 &  1.8$\pm$0.8 &    90\% &  $1.8\pm 0.5$ &  \citet{rampazzo07} \\
4382 &  1.8$\pm$0.8 &    85\% &  $1.6\pm 0.3$ &  \citet{sansom06}   \\
7727 &  0.9$\pm$0.8 &   100\% &  1\,--\,2     &  \citet{trancho04}
\enddata

\tablenotetext{a}{Age of young metal-rich GC subpopulation as determined
from \chisq\ contours.}
\tablenotetext{b}{Percentage of observed GC candidates belonging to the
younger subpopulation as determined from \chisq\ contours.}
\label{table4}
\end{deluxetable*}

\subsubsection{\n4382}

For \n4382, the \chisq\ contours shown in Figure~\ref{fig:chi2_contour}
indicate a younger metal-rich GC subpopulation of age $1.8\pm 0.8$ Gyr,
comprising between about 75\% and 95\% of the observed GC candidates.

Note that the \chisq\ values for the GCs in \n4382 are much smaller than
those for the GCs in the other three galaxies, suggesting a better agreement
between the observed GC data and the simulated cluster populations.

As can be seen, there is also a significantly wider spread for the ratio of
young-to-old GCs.
Recall, however, that our primary interest is in the age of the younger
subpopulation, and note that the convergence in age is very high.
This newly determined age for the younger metal-rich GC subpopulation agrees
well with the previously published age of $1.6 \pm 0.3$ Gyr for the young
{\em stellar} subpopulation of this galaxy \citep{sansom06}.

\subsubsection{\n7727}

Finally, for \n7727 the \chisq\ contours of Figure~\ref{fig:chi2_contour}
indicate a purely young metal-rich GC population of age $0.9 \pm 0.8$ Gyr.
The observed GC candidates of this galaxy are noteworthy both for their
particularly young ages and for the apparent absence of an older
subpopulation.
Both of these characteristics were previously { reported} by
\citet{trancho04}, who obtained a GC population age of 1\,--\,2 Gyrs. 
Note, however, that the population of old metal-poor GCs universally present
in all major galaxies may be too faint to be detected on our \ks-band images.
Hence, its apparent absence { may simply reflect a selection effect and
does not imply that this old metal-poor GC population doesn't exist.}

\section{Summary and Conclusions}\label{sec:disc}

We have presented an age analysis of GC systems in four early-type galaxies
that are considered candidate remnants of recent mergers.
New \ks-band photometry of the GCs in these galaxies has been obtained with
two near-infrared imagers on Gemini and combined with existing \hst\,
photometry in $V$ and $I$ to produce color--color and color--magnitude
diagrams of the GC systems in \viks.
These data have been fitted to simple toy models of mixed GC populations
containing three subpopulations that differ in age and metallicity.
The mixed GC populations have been simulated with SSP models by Charlot \&
Bruzual (2007, private communication) and compared to the observations.
Various \chisq\ fits to the cumulative age distributions derived from our
data and from the toy models yield clear evidence for the presence of
intermediate-age, metal-rich subpopulations of GCs in each of the
four early-type galaxies.
At the observed, relatively bright \ks\ magnitudes these intermediate-age
subpopulations exceed the old metal-rich and metal-poor subpopulations in
cluster numbers, and their estimated ages of $\sim$\,1\,--\,2 Gyr are
consistent with published ages for the same galaxies based on the integrated
light.
Thus, \viks\ photometry of GC populations in early-type galaxies can be used
to date epochs of major star and cluster formation presumably due to
relatively recent merger events.

\begin{acknowledgements}
We thank St\'ephane Charlot and Gustavo Bruzual for sending us an early
version of their new cluster-population models (``CB07''), presently still
under development, and Maren Hempel and Markus Kissler-Patig for helpful
information and a copy of their simulation code.
We also thank the anonymous referee, whose constructive comments helped
us improve the presentation of our results.
This research is based on observations obtained at the Gemini Observatory,
which is operated
by the Association of Universities for Research in Astronomy, Inc., under a
cooperative agreement with the NSF on behalf of the Gemini partnership: the
National Science Foundation (United States), the Particle Physics and
Astronomy Research Council (United Kingdom), the National Research Council
(Canada), CONICYT (Chile), the Australian Research Council (Australia), CNPq
(Brazil) and CONICET (Argentina).
\end{acknowledgements}

%%%%%%%%%%%%%%%%%%%%%%%%%%%%%%%   REFERENCES   %%%%%%%%%%%%%%%%%%%%%%%%%%%%%%%%

\end{document}